# SmI$_3$: 4$f^5$ honeycomb magnet with spin-orbital entangled Γ$_7$ Kramers doublet

Hajime Ishikawa[1], Ryosuke Kurihara[1], Takeshi Yajima[1], Daisuke Nishio-Hamane[1], Yusei Shimizu[2], Toshiro Sakakibara[1], Akira Matsuo[1], Koichi Kindo[1]

[1]Institute for Solid State Physics, the University of Tokyo, 5-1-5 Kashiwanoha, Kashiwa, Chiba, Japan
[2]Institute for Materials Research, Tohoku University, Oarai, Ibaraki, 311-1313, Japan

E-mail: hishikawa@issp.u-tokyo.ac.jp

We report magnetic properties of a 4$f$-honeycomb iodide SmI$_3$ made up of edge-shared network of SmI$_6$ octahedra. High temperature magnetic susceptibility indicates Γ$_7$ Kramers doublet ground state of Sm$^{3+}$ (4$f^5$) ions stabilized by the spin-orbit coupling and octahedral crystal electric field, which interact with Sm-I-Sm bond angle nearly 90°. Magnetization measurements down to 0.1 K detected antiferromagnetic correlations and an anomaly in the magnetization curve before saturation without a sign of long-range order. Relevance between SmI$_3$ and the antiferromagnetic Kitaev material proposed in the 4$f$-electron system is discussed.

## I. INTRODUCTION

Quantum spin liquid (QSL) is an exotic electronic state of matter where long-range magnetic order is absent in spite of the presence of magnetic correlations among spins [1,2] and has attracted condensed matter physicists since the proposal of resonating valence bonds by Anderson [3]. Realization of QSL has been mainly discussed in geometrically frustrated spin systems such as kagome and pyrochlore antiferromagnets [4,5]. In recent years, a theoretical model which exhibits a well-defined QLS ground state, so-called the Kitaev model, emerged as a platform for QSL physics [6]. Theoretical investigations revealed that a spin-1/2 model on honeycomb lattice with bond-dependent Ising-type, which is often called Kitaev-type, interaction exhibits a QSL ground state where fractional excitations emerge [6,7,8,9].

Kitaev-type interaction is proposed to appear in the Mott insulator made of magnetic ions with spin-orbital entangled Kramers doublet ground state [10,11]. When the ligand octahedra containing the magnetic metal ions are connected via edges with metal-ligand-metal bond angle of 90°, Kitaev-type superexchange interaction may emerge while the isotropic Heisenberg-type superexchange interaction is suppressed [10,11]. As candidate materials that fulfill these conditions, layered oxides and chlorides with $Ir^{4+}$ ($5d^5$) [12, 13, 14, 15] and $Ru^{3+}$ ($4d^5$) [16, 17] ions have been investigated: see also reviews [7,8].

4$f$-electron systems attract attention as largely unexplored target materials for investigating the Kitaev spin liquid physics in recent years [11]. Kitaev-type interaction is proposed to appear between 4$f$-ions with Kramers doublet ground state stabilized by spin-orbit coupling and the octahedral crystal electric field (CEF) such as $\Gamma_7$ state of the 4$f^1$ ion [11,18]. A major difference between the $d$- and $f$- electron cases is the sign of Kitaev-type interaction: ferromagnetic and antiferromagnetic interactions are expected in the $d$- and the $f$-electron cases, respectively [11,18]. While the QSL state appears for both cases at zero magnetic field [7,19], antiferromagnetic Kitaev model may allow us to investigate another magnetic-field-induced QSL phase in a magnetic field range before spins are polarized [19,20,21]. Materials containing the honeycomb lattice of $Pr^{4+}$ ($4f^1$) [22,23] and $Yb^{3+}$ ($4f^{13}$) [24,25,26,27] are investigated as candidates, however, a 4$f$-honeycomb material with well-defined $\Gamma_7$ state is still missing.

We focus on samarium triiodide ($SmI_3$) with the $BiI_3$-type structure (Fig.1(a)) [28], of which magnetic properties are not known. The spin-orbit coupling of 1050 cm$^{-1}$ of $Sm^{3+}$ ($4f^5$) ion with $L$ = 5 and $S$ = 5/2 [29], which corresponds to approximately 1500 K, would stabilize the $J$ = 5/2 sextet at ambient condition. The degeneracy of the sextet can be lifted by the octahedral CEF to stabilize the $\Gamma_7$ Kramers doublet (Fig.1(a)) [11]. The edge-shared network of $SmI_6$ octahedra forming a honeycomb lattice materializes the situation discussed in the theoretical study [11,18]. We synthesized the powder and single crystal samples of $SmI_3$ and investigated the magnetic properties. Magnetic susceptibility at high temperature indicates the $\Gamma_7$ Kramers doublet state of the $Sm^{3+}$ ion stabilized by the CEF splitting approximately 200 K. The magnetization measured down to 0.1 K detected the development of antiferromagnetic correlations without a sign of long-range magnetic order. Magnetization curve at 0.1 K exhibits an anomaly at around 2.5 T before polarized state is reached at around 5 T. We argue $SmI_3$ is a rare 4$f$-electron system that hosts well-defined $\Gamma_7$ Kramers doublet ions interacting on the honeycomb lattice made of edge-shared metal-ligand octahedra.

## II. SAMPLE PREPARATION AND CHARACTERIZATION

Polycrystalline powder of $SmI_3$ is obtained by reacting samarium powder (99.9%) and iodine (99.999%) at 650°C in the evacuated quartz glass tube. To prevent possible formation of $SmI_2$, excess iodine is added with the molar ratio of Sm : I = 1 : 6. Excess iodine is removed by sublimation after the reaction. All the materials were handled in the argon-filled glove box. Powder x-ray diffraction measurement by the diffractometer with Cu-K$\alpha_1$ radiation (RIGAKU, Smart Lab) indicates the BiI$_3$-type structure (R-3) with the lattice constants $a$ = 7.63512(3) Å and $c$ = 20.8712(1) Å at 250 K (Fig1(b), depicted by VESTA software [30]). Extra peaks in the powder pattern are attributed to Apiezon-N grease used to protect the air-sensitive sample and tiny amount of SmOI impurity (Fig.1(d)). Atomic positions determined from the Rietveld analysis of the powder pattern using the FullProf software [31] revealed the Sm-I-Sm bond angle of 90.5(1)°.

To obtain single crystals, polycrystalline sample was sublimated in a tube furnace at around 800°C with a temperature gradient. Hexagonal shaped crystals up to 5 mm were obtained (Fig.1(c)). Chemical analysis was performed using scanning electron microscopy (SEM, JEOL IT-100) equipped with energy dispersive X-ray spectroscopy (EDS, 15 kV, 0.8 nA, 10 $\mu$m beam diameter) at the Institute for Solid State Physics (ISSP), the university of Tokyo. The data correction was performed by the ZAF method which takes into account atomic number (Z) effect, absorption (A) effect, and fluorescence excitation (F) effect. The analysis revealed the chemical composition of Sm : I = 1 : 3: see supplemental material for detail [32]. In the x-ray diffraction pattern of a hexagonal crystal put parallel to the sample holder, strong 00$l$ ($l$ = 3$n$, $n$ = interger) diffractions are observed (Fig.1(d)), indicating the hexagonal plane corresponds to the honeycomb layer.

### III. MAGNETIC PROPERTY

Temperature dependence of magnetic susceptibility $\chi$ between 2 and 300 K was measured by the SQUID magnetometer (MPMS-XL, Quantum Design). $\chi$ of the powder, a bunch of sublimated small crystals typically smaller than 1 mm and a pile of large single crystals exhibit nearly identical behavior at high temperature (Fig.2(a)): data for different samples are shown with offsets. Curie-Weiss like behavior is observed above 100 K and a shoulder is observed at around 50 K. $\chi$ exhibits an upturn toward low temperature below 30 K. $\chi$ of a pile of large crystals was measured in magnetic field applied perpendicular and parallel to the hexagonal plane, which correspond to $H$ // $c$ and $H \perp c$, respectively. The in-plane orientation of the crystals was not aligned. $\chi$ in $H \perp c$ exhibits a more pronounced shoulder at 50 K. The upturn at low temperature is enhanced in $H$ // $c$ compared to $H \perp c$.

In order to elucidate $\chi$, the magnetic susceptibility of Sm$^{3+}$ ion ($\chi_{ion}$) under cubic octahedral CEF Hamiltonian, $H_{CEF} = B_4^0(O_4^0 + 5O_4^4)$, which splits the $J$ = 5/2 sextet into the $\Gamma_7$ doublet and $\Gamma_8$ quartet, was calculated. We calculated $\chi$ by considering the external magnetic field as perturbation as shown in Ref. [33]. The matrix elements used in the calculation are shown in the supplemental material [32]. The calculations were performed using the Julia language (ver. 1.2.0). In the calculation, two contributions that are often called Curie- and van Vleck- terms appear in $\chi$. The Curie-term diverges toward low temperature, while the van Vleck-term takes a constant value at low temperature. The van Vleck-term produces a shoulder in $\chi$ at the temperature determined by the size of the CEF splitting. The prominence of the shoulder depends on the relative magnitude the two terms, which is determined by the type of the ground state wave function. The shoulder is prominent for the $\Gamma_7$ ground state but is negligibly small for the $\Gamma_8$ ground state: see supplemental material for detail [32]. The observation of the clear shoulder in $\chi$ indicates the $Sm^{3+}$ ions host the $\Gamma_7$ doublet ground state in $SmI_3$. This is reasonable as the Coulomb repulsion between 4$f$-electron and $I^-$ ions is smaller for the $\Gamma_7$ state compared to the $\Gamma_8$ state in the octahedral coordination.

To estimate the size of the CEF splitting, $\chi$ was calculated by changing the value of $B_4^0$. The calculations revealed that characteristic shoulder around 50 K is best reproduced with the CEF splitting of 216 K ($B_4^0$ = 0.6 K): see supplemental material for the results with different values of $B_4^0$ [32]. The molar susceptibility $\chi_{mol}$ is calculated as $\chi_{mol} = N_A(g\mu_B)^2\chi_{ion}/k_B + \chi_0$, where $N_A$, $g$, $\mu_B$, $k_B$, and $\chi_0$ indicate Avogadro constant, $g$-factor, Bohr magneton, Boltzmann constant, and a temperature independent term, respectively, and compared with experimental results in $H // c$ ((Fig.2(a)). A calculation using the Landé $g$-factor of 2/7 of the $Sm^{3+}$ (4$f^5$) ion ($J$ = 5/2, $L$ = 5 and $S$ = 5/2) and $\chi_0$ = 5.5 × $10^{-4}$ emu/mol, which can be attributed to the contribution from higher energy $J$ = 7/2 multiplet, yields a good fit of the high-temperature experimental data. Note that the shoulder is almost invisible in the $\chi$ calculated using the same parameter set for the $\Gamma_8$ ground state (Fig.2(a)).

Magnetization curve of sublimated small crystals was measured by the induction method using a pick-up coil in the pulsed magnetic field up to around 42 T generated at the International MegaGauss Science Laboratory at ISSP. The magnetization curve is nearly linear at 4.2 K in the whole magnetic field range (Fig.2(a)). The curve exhibits a convex shape at 1.4 and 0.8 K below 20 T and increases linearly at higher fields. The featureless magnetization curve suggests the paramagnetic state at 0.8 K.

Specific heat of $SmI_3$ was measured for a small pellet made from sublimated small crystals down to 0.6 K by the relaxation method using a commercial apparatus (PPMS, Quantum Design). The specific heat divided by temperature $C_p/T$ exhibits a shoulder around 15 K. Similar shoulder was observed in the nonmagnetic and isostructural $BiI_3$ (99.9%), indicating the shoulder is attributed to the lattice contribution. $C_p/T$ of $SmI_3$ increases below 2 K as lowering the temperature. As the lattice contribution is small in the temperature range, the upturn should be of magnetic origin. The lattice specific heat below 5 K is estimated by assuming $C \propto T^3$ according to the Debye model and the leftover magnetic contribution is estimated (Fig.2(b)). The magnetic entropy $S_m$ obtained by integrating the magnetic contribution between 0.6 and 5 K amounts to 0.42 J/mol-K and is approximately 7 % of $R\ln 2$ expected for the $\Gamma_7$ doublet. The estimation suggests the rest of the magnetic entropy should be released below 0.6 K.

Magnetization of a pile of large single crystals was measured below 2 K in $H // c$ down to 0.1 K in a dilution refrigerator by a capacitive Faraday method [34,35] (Fig.3(a)). $\chi$ exhibits a monotonic behavior without a clear kink or drop suggestive of the long-range antiferromagnetic order down to 0.1 K. Note that $\chi$ measured at weak magnetic fields exhibits larger values, which can be attributed to the presence of a tiny amount of ferromagnetic impurity: see supplemental material for detail [32]. When the impurity contribution is subtracted, $\chi$ at different magnetic fields almost overlap with each other (inset in Fig.3(a)). $\chi$ below 10 K is slightly larger than the calculation for the free ion (dashed line in Fig.3(a)) but suppressed below 2 K as lowering the temperature, suggesting the presence of antiferromagnetic correlations. The effect of antiferromagnetic correlations is prominent below 0.2 K where $\chi$ exhibits a broad maximum.

Magnetization curve of the single crystal sample is measured at 0.1 K up to 8 T (Fig.3(b)). The curve exhibits a clear kink at around 5 T and increases gradually at higher fields, suggesting the saturation of the magnetic moment of the $\Gamma_7$ doublet. The calculated magnetization curve of a free ion at 0.1 K exhibits the saturation behavior at around 1 T (solid line in Fig.3 (b)). The shift of the magnetic saturation to higher magnetic field is consistent with the presence of an antiferromagnetic correlations. Moreover, the magnetic field derivative of magnetization $dM/dH$ at 0.1 K exhibits a broad peak at around 2.5 T. Such an anomaly is absent in the calculation for the free ion and is smeared out above 0.2 K.

## IV. DISCUSSION

CEF splitting of approximately 200 K estimated from the high temperature magnetic susceptibility is much smaller compared to the energy scale of the spin-orbit coupling of the $Sm^{3+}$ ion ~ 1500 K [29], supporting the validity of the energy splitting diagram shown in Fig.1(a). Theoretical studies point out that antiferromagnetic Kitaev-type superexchange interaction may emerge when the 4$f$ ions with the $\Gamma_7$ ground state interact with the metal-ligand-metal angle of 90° [11,18]. These conditions are perfectly fulfilled in $SmI_3$.

Search for the Kitaev material candidate in 4f-electron systems have carried out on the honeycomb materials with $Pr^{4+}$ ($4f^1$) or $Yb^{3+}$ ($4f^{13}$) ions, where the $\Gamma_7$ doublet can be the ground state [11]. Neutron scattering study on a $4f^1$ honeycomb oxide $Na_2PrO_3$ detected the first excited state at 0.23 eV [23]. This energy is larger than the spin-orbit coupling of the $Pr^{4+}$ ion ~ 0.12 eV [18] and it is suggested that a simple $\Gamma_7$ picture is not appropriate [23]. In the case of $Yb^{3+}$ ($4f^{13}$) ion seen in the honeycomb compounds $YbCl_3$ and $YbBr_3$ [24,25,26,27], the octahedral CEF may stabilize $\Gamma_6$ state rather than the $\Gamma_7$ state and the dominant interaction can be the Heisenberg-type [25]. Neutron scattering experiments revealed that the dominant exchange interaction is Heisenberg-type [26,27]. $SmI_3$ is regarded as a rare example that hosts well-defined $\Gamma_7$ state of the 4f-ion on the honeycomb lattice.

Superexchange interactions between magnetic ions with $\Gamma_7$ states emerging from the $J = 5/2$ sextet were investigated in detail on the $Pr^{4+}$ ($4f^1$) oxides [18]. In the case of the $4f^1$ ion, the $J = 5/2$ sextet is formed by the coupling of $L = 3$ and $S = 1/2$. While the $J = 5/2$ sextet is formed by the coupling of $L = 5$ and $S = 5/2$ in the case of $4f^5$ ion, the wave function takes the identical form with the $4f^1$ ion [11]. Our results call for theoretical study on how the number of electron and the magnetic cation and ligand anion species alter the superexchange interactions between the $\Gamma_7$ ions.

We compare $SmI_3$ and $4/5d^5$ honeycomb magnets, which may host the $\Gamma_7$ doublet ground state arising from the coupling of $L_{\text{eff}} = 1$ and $S = 1/2$ and are extensively investigated as materials with ferromagnetic Kitaev-type interaction. From the onset of the release of magnetic entropy and the deviation of $\chi$ from the free ion values, the energy scale of magnetic interactions in $SmI_3$ should be of the order of 1 K. This is 10-100 times smaller compared to the d-electron materials [7,8]. Well-investigated $A_2IrO_3$ and $RuCl_3$ exhibit a long-range magnetic order at 15 and 7 K, respectively, which are ~ 1/10 of the exchange interactions [7,8]. QSL ground state is achieved in iridates when the intercalant cations are replaced by other cations via ion-exchange reaction from parent $A_2IrO_3$ [13,14,15]. It is often difficult to obtain a large single crystal via ion-exchange reaction and the effect of randomness in the ion-exchanged compounds on magnetism is not fully understood [7]. While the small exchange interaction in $SmI_3$ requires very low temperature experiments, the absence of the magnetic order down to 0.1 K and the availability of single crystal without the interlayer cation would make $SmI_3$ a promising target for further experimental investigations.

We comment on the broad peak in $dM/dH$ at 0.1 K at around 2.5 T, which is suggestive of a change in the magnetic state. The peak does not appear in the calculation for the free ion and appears only below 0.2 K, indicating the antiferromagnetic correlations play a crucial role. In the theoretical study on the antiferromagnetic Kitaev model, a magnetic-field-induced transition from the Kitaev spin liquid to another QSL state is proposed at the intermediate field region before polarized state is achieved [19,20,21]. It would be an interesting future work to examine whether the anomaly in the magnetization curve is relevant to the theoretical prediction.

## V. CONCLUSION AND PERSPECTIVE

In summary, magnetic properties of the layered honeycomb iodide $SmI_3$ are investigated. We demonstrate $Sm^{3+}$ ($4f^5$) ions arranged on the honeycomb lattice host the $\Gamma_7$ Kramers doublet ground state and interact under the edge-shared geometry of $SmI_6$ octahedra. Development of the antiferromagnetic correlations and an anomaly in the magnetization process are detected, while clear signature of the long-range order is not observed down to 0.1 K. $SmI_3$ extends the family of spin-orbital entangled honeycomb magnet and would deserve further investigations as a rare 4$f$-honeycomb system with $\Gamma_7$ Kramers doublet state. Specific heat measurements and torque magnetometry at dilution refrigerator temperatures would be useful for investigating the magnetic state and magnetic anisotropy. Spectroscopic measurements using muon and neutron would also be useful to clarify the magnetic state and the nature of the magnetic interactions, while the sample prepared from isotopically enriched Sm is necessary for the neutron experiments to reduce neutron absorption. Our results call for theoretical investigation on the superexchange interaction between $Sm^{3+}$ ($4f^5$) ions and would trigger the explorations of materials comprising the honeycomb lattice of $Sm^{3+}$ ions.

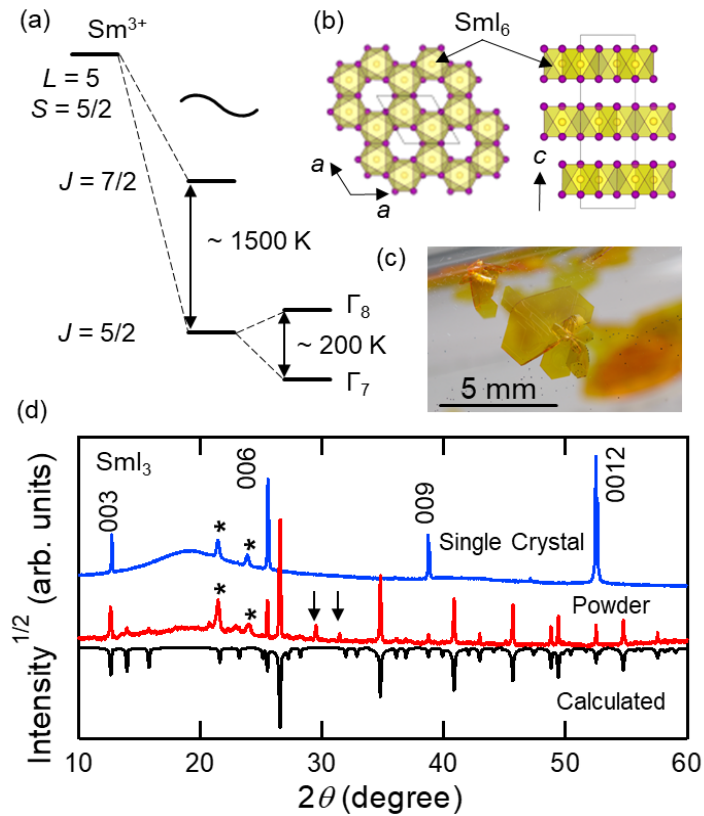

Figure 1

(a) Energy diagram of $Sm^{3+}$ ($4f^5$) ion in $SmI_3$. (b) A honeycomb layer in $SmI_3$ seen from *c*-axis (left) and the crystal structure seen from *a*-axis (right). (c) Picture of the single crystals of $SmI_3$ inside the quartz glass tube. (d) X-ray diffraction pattern of the powder (red) and the single crystal (blue) of $SmI_3$ compared with the calculated pattern (black). The broad feature around $2\theta = 20°$ and the peaks indicated by * come from the grease used to protect the sample. The peaks indicated by the arrows in the powder sample come from the SmOI impurity.

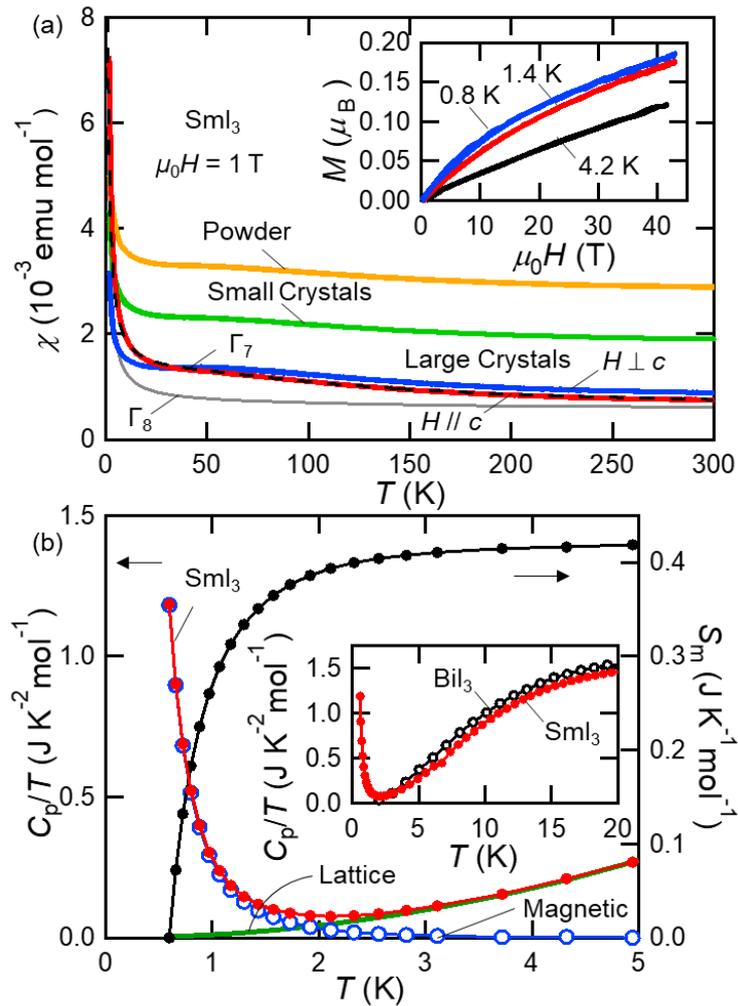

Figure 2

(a) Magnetic susceptibility of $SmI_3$. The data for powder and a bunch of small crystals are shown with offsets. The magnetization curves of sublimated small crystals are shown in the inset. (b) Specific heat divided by temperature and electronic entropy of sublimated small crystals of $SmI_3$. Data below 20 K is shown in the inset and compared with the nonmagnetic analogue $BiI_3$. Estimated lattice and magnetic contributions are shown by solid line and unfilled markers, respectively.

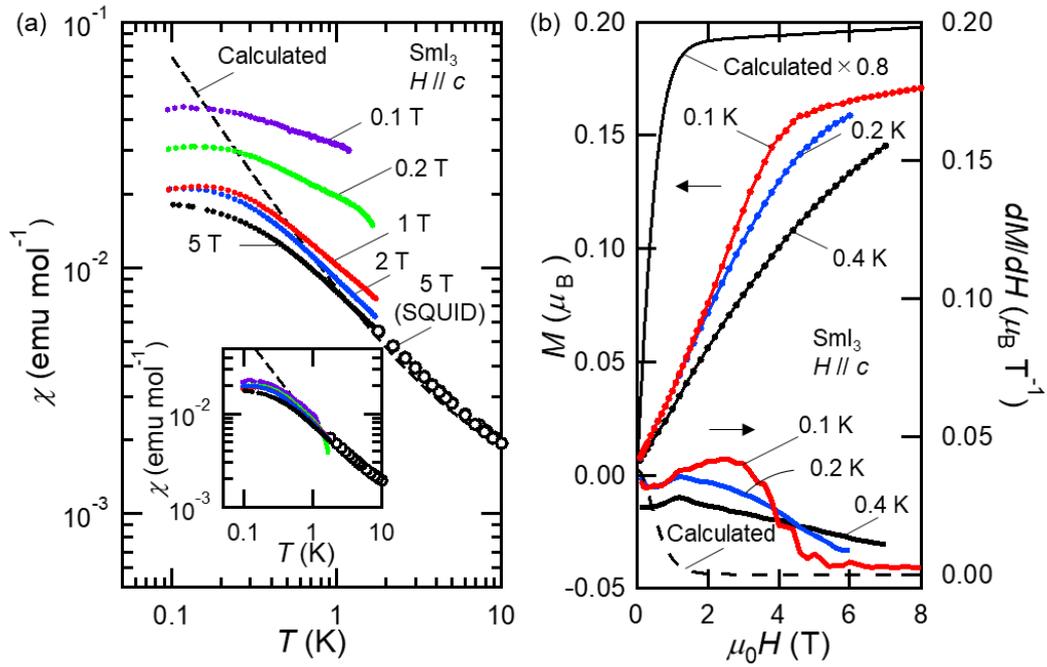

Figure 3

(a) Magnetic susceptibility of a pile of single crystals of $SmI_3$ measured down to 0.1 K in $H \parallel c$ at different magnetic fields. The same data after subtracting the impurity component estimated as in [32] are shown in the inset. (b) Magnetization curve of $SmI_3$ in $H \parallel c$ and its derivative by magnetic field. The solid line indicates the calculation for the free ion multiplied by a factor of 0.8.

Supplemental Material for

# SmI$_3$: 4$f^5$ honeycomb magnet with spin-orbital entangled Γ$_7$ Kramers doublet

Hajime Ishikawa[1], Ryosuke Kurihara[1], Takeshi Yajima[1], Daisuke Nishio-Hamane[1],
Yusei Shimizu[2], Toshiro Sakakibara[1], Akira Matsuo[1], Koichi Kindo[1]

[1]Institute for Solid State Physics, the University of Tokyo, 5-1-5 Kashiwanoha, Kashiwa, Chiba, Japan
[2]Institute for Materials Research, Tohoku University, Oarai, Ibaraki, 311-1313, Japan

1. Rietveld refinement of the powder x-ray diffraction pattern of SmI$_3$

Powder x-ray diffraction measurement was performed by the diffractometer with Cu-K$\alpha_1$ radiation (RIGAKU, Smart Lab) at 250 K. Rietveld analysis of the powder pattern was performed using the FullProf software [S1]. Date below 2$\theta$ = 25° with large backgrounds from Apiezon-N grease used to protect the air-sensitive sample and peaks from SmOI impurity around 2$\theta$ = 30° are avoided in the refinement. The observed and calculated powder patterns are shown by the red plots and black line, respectively. The difference between the observed and calculated intensities are shown by the blue line with an offset.

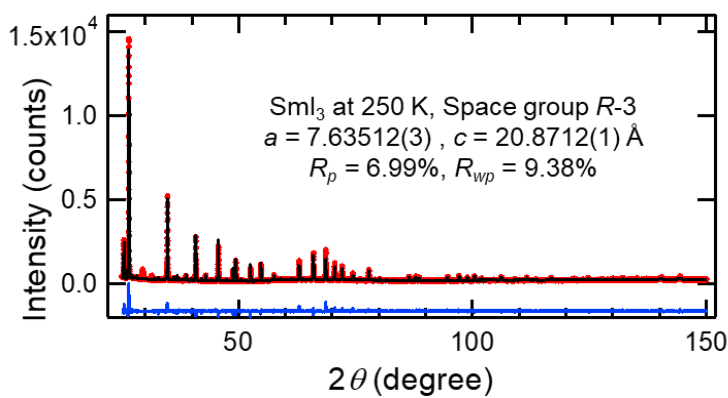

Atomic Positions

|    | x         | y         | z         | Occupancy | $B_{iso}$ |
|----|-----------|-----------|-----------|-----------|-----------|
| Sm | 0         | 0         | 0.1663(2) | 1         | 1.42      |
| I  | 0.3333(5) | 0.3337(5) | 0.0817(1) | 1         | 1.22      |

2. Elemental analysis of the single crystals of SmI$_3$

Chemical analysis was performed for the sublimated large crystals of SmI$_3$ using scanning electron microscopy (SEM, JEOL IT-100) equipped with energy dispersive X-ray spectroscopy (EDS, 15 kV, 0.8 nA, 10 μm beam diameter). The data correction was performed by the ZAF method which takes into account atomic number (Z) effect, absorption (A) effect, and fluorescence excitation (F) effect [S2]. The chemical compositions at selected points marked by the blue square in the left pictures of the crystal (left) are shown in the table (right).

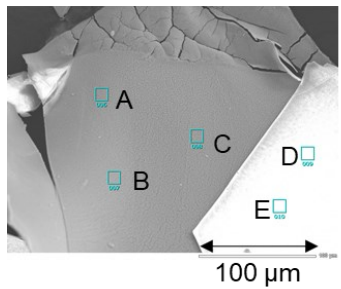 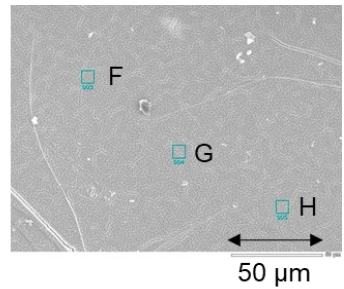

|   | Sm   | I    |
|---|------|------|
| A | 0.99 | 3.01 |
| B | 1.00 | 3.00 |
| C | 0.98 | 3.02 |
| D | 0.94 | 3.06 |
| E | 0.95 | 3.05 |
| F | 1.06 | 2.94 |
| G | 1.01 | 2.99 |
| H | 0.97 | 3.03 |

3. Estimation of the magnetic impurity in $SmI_3$ at low temperature

Small and positive offsets are seen in the magnetization curves measured at low temperature as shown below, suggesting the presence of a tiny amount of ferromagnetic impurity. In the data shown in the inset of Fig.3(a), corresponding amount of constant magnetization is subtracted from the magnetic susceptibility data shown in the main panel. The data at 0.1 T and 0.2 T exhibit good agreement with the data at higher magnetic fields. The data obtained at higher fields is almost unaffected due to the larger signals compared to the offset.

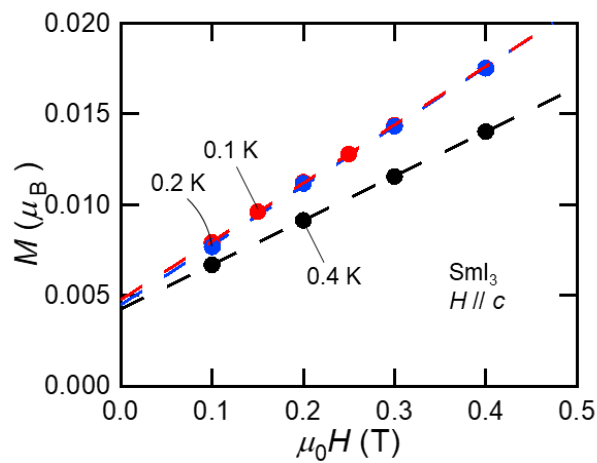

## 4. Calculation of magnetic susceptibility

We calculated magnetic susceptibility $\chi$ by considering the external magnetic field as perturbation following the reference [S3]. In the calculation, two contributions that are often called Curie- and van Vleck- terms appear in $\chi$. The Curie-term ($\chi_{Curie}$) diverges toward low temperature, while the van Vleck-term ($\chi_{Van\ Vlek}$) takes a constant value at low temperature. $\chi_{van\ Vlek}$ produces a shoulder in the sum of the two terms ($\chi_{Total}$) at the temperature determined by the size of the crystal electric field (CEF) splitting. Comparing the $\chi$ for the same size of the CEF splitting ($\Delta_{CEF}$), the shoulder is prominent in the case of $\Gamma_7$ ground state compared to the $\Gamma_8$ ground state due to the difference in relative magnitude of $\chi_{Curie}$ and $\chi_{Van\ Vlek}$ (a,b).

In order to estimate $\Delta_{CEF}$, we performed calculations by changing the value of $\Delta_{CEF}$. The calculation with $\Delta_{CEF}$ = 216 K well reproduces the experimentally observed shoulder (c). Smaller (larger) $\Delta_{CEF}$ enhances (suppresses) the $\chi_{Van\ Vlek}$ at low-temperature region and shift the shoulder to lower (higher) temperature, yielding the worse fitting.

The striking difference between the $\chi$ for the $\Gamma_7$ and $\Gamma_8$ ground states comes from the difference in the ground state wave function: the wave functions and matrix elements used in the calculation are shown below (d). Note that the conventional cubic axis is used. To compare the experimentally obtained $\chi$ of SmI$_3$ in $H$ // $c$ and the calculation, we calculated the $\chi$ in magnetic field along [111] direction by considering $J_{[111]} = (J_x + J_y + J_z)/\sqrt{3}$.

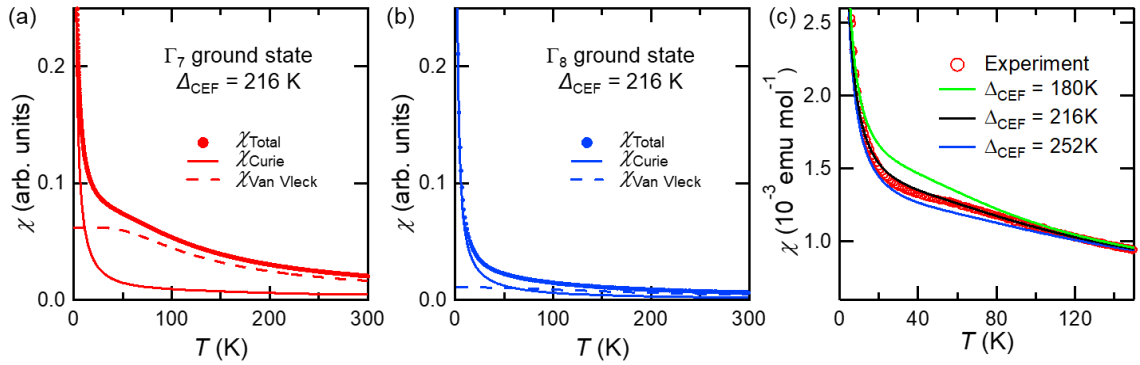

(d)

$$J_x = \begin{pmatrix} & |\Gamma_8^{1-}\rangle & |\Gamma_8^{1+}\rangle & |\Gamma_8^{2-}\rangle & |\Gamma_8^{2+}\rangle & |\Gamma_7^{1-}\rangle & |\Gamma_7^{1+}\rangle \\ \langle\Gamma_8^{1-}| & 0 & \frac{3}{2} & & \sqrt{\frac{1}{3}} & & -\sqrt{\frac{5}{3}} \\ \langle\Gamma_8^{1+}| & \frac{3}{2} & 0 & \sqrt{\frac{1}{3}} & & -\sqrt{\frac{5}{3}} & \\ \langle\Gamma_8^{2-}| & & \sqrt{\frac{1}{3}} & 0 & \frac{5}{6} & & -\frac{\sqrt{5}}{3} \\ \langle\Gamma_8^{2+}| & \sqrt{\frac{1}{3}} & & \frac{5}{6} & 0 & -\frac{\sqrt{5}}{3} & \\ \langle\Gamma_7^{1-}| & & -\sqrt{\frac{5}{3}} & & -\frac{\sqrt{5}}{3} & 0 & -\frac{5}{6} \\ \langle\Gamma_7^{1+}| & -\sqrt{\frac{5}{3}} & & -\frac{\sqrt{5}}{3} & & -\frac{5}{6} & 0 \end{pmatrix}$$

$$J_y = -i\begin{pmatrix} & |\Gamma_8^{1-}\rangle & |\Gamma_8^{1+}\rangle & |\Gamma_8^{2-}\rangle & |\Gamma_8^{2+}\rangle & |\Gamma_7^{1-}\rangle & |\Gamma_7^{1+}\rangle \\ \langle\Gamma_8^{1-}| & 0 & -\frac{3}{2} & & \sqrt{\frac{1}{3}} & & -\sqrt{\frac{5}{3}} \\ \langle\Gamma_8^{1+}| & \frac{3}{2} & 0 & -\sqrt{\frac{1}{3}} & & \sqrt{\frac{5}{3}} & \\ \langle\Gamma_8^{2-}| & & \sqrt{\frac{1}{3}} & 0 & -\frac{5}{6} & & \frac{\sqrt{5}}{3} \\ \langle\Gamma_8^{2+}| & -\sqrt{\frac{1}{3}} & & \frac{5}{6} & 0 & -\frac{\sqrt{5}}{3} & \\ \langle\Gamma_7^{1-}| & & -\sqrt{\frac{5}{3}} & & \frac{\sqrt{5}}{3} & 0 & \frac{5}{6} \\ \langle\Gamma_7^{1+}| & \sqrt{\frac{5}{3}} & & -\frac{\sqrt{5}}{3} & & -\frac{5}{6} & 0 \end{pmatrix}$$

$$J_z = \begin{pmatrix} & |\Gamma_8^{1-}\rangle & |\Gamma_8^{1+}\rangle & |\Gamma_8^{2-}\rangle & |\Gamma_8^{2+}\rangle & |\Gamma_7^{1-}\rangle & |\Gamma_7^{1+}\rangle \\ \langle\Gamma_8^{1-}| & -\frac{1}{2} & & & & & \\ \langle\Gamma_8^{1+}| & & \frac{1}{2} & & & & \\ \langle\Gamma_8^{2-}| & & & -\frac{11}{6} & & -\frac{2\sqrt{5}}{3} & \\ \langle\Gamma_8^{2+}| & & & & \frac{11}{6} & & \frac{2\sqrt{5}}{3} \\ \langle\Gamma_7^{1-}| & & & -\frac{2\sqrt{5}}{3} & & \frac{5}{6} & \\ \langle\Gamma_7^{1+}| & & & & \frac{2\sqrt{5}}{3} & & -\frac{5}{6} \end{pmatrix}$$

$$\left|\Gamma_8^{1-}\right\rangle = \left|-\frac{1}{2}\right\rangle$$

$$\left|\Gamma_8^{1+}\right\rangle = \left|+\frac{1}{2}\right\rangle$$

$$\left|\Gamma_7^{1-}\right\rangle = \sqrt{\frac{1}{6}}\left|-\frac{5}{2}\right\rangle - \sqrt{\frac{5}{6}}\left|+\frac{3}{2}\right\rangle$$

$$\left|\Gamma_8^{2-}\right\rangle = \sqrt{\frac{5}{6}}\left|-\frac{5}{2}\right\rangle + \sqrt{\frac{1}{6}}\left|+\frac{3}{2}\right\rangle$$

$$\left|\Gamma_7^{1+}\right\rangle = \sqrt{\frac{1}{6}}\left|+\frac{5}{2}\right\rangle - \sqrt{\frac{5}{6}}\left|-\frac{3}{2}\right\rangle$$

$$\left|\Gamma_8^{2+}\right\rangle = \sqrt{\frac{1}{6}}\left|+\frac{5}{2}\right\rangle + \sqrt{\frac{5}{6}}\left|-\frac{3}{2}\right\rangle$$

$$J_{[111]} = \frac{J_x + J_y + J_z}{\sqrt{3}}$$